\documentclass[12pt,preprint]{aastex}
\usepackage[onecolumn]{emulateapj5}  

\newcommand{\apn}{\alpha_{0.3}}         
\newcommand{\betas}{\beta_*}            
\newcommand{\cs}{c_{\rm s}}             
\newcommand{\epn}{\varepsilon_{0.1}}    
\newcommand{\krel}{\hat\kappa}          
\newcommand{\kB}{k_{\rm B}}             
\newcommand{\lum}{l_{\scriptscriptstyle\rm E}} 

\newcommand{\mH}{m_{\rm H}}

\newcommand{\pgas}{p_{\rm gas}}
\newcommand{\prad}{p_{\rm rad}}
\newcommand{\rh}{R_{H}}

\newcommand{\rs}{R_{\rm S}}
\newcommand{\rsg}{r_{\rm s.g.}}

\newcommand{\Teff}{T_{\rm eff}}
\newcommand{\tkh}{t_{\rm KH}}

\newcommand{\veps}{\varepsilon}

\newcommand{\unit}[1]{\,{\rm #1}}
\newcommand{\au}{\,\mbox{\textsc{au}}}
\newcommand{\cm}{\unit{cm}}

\newcommand{\ergs}{\unit{erg}\unit{s}^{-1}}
\newcommand{\g}{\unit{g}}
\newcommand{\K}{\unit{K}}

\newcommand{\kms}{\unit{km\,s^{-1}}}

\newcommand{\msun}{M_\odot}
\newcommand{\pc}{\unit{pc}}

\newcommand{\yr}{\unit{yr}}

\newcommand{\eqref}[1]{eq.~(\ref{#1})}

\newcommand{\upi}{\pi}

\newcommand{\smyr}{{ M_\odot\ \rm yr^{-1}}}
\newcommand{\sm}{{ M_\odot}}

\newcommand{\beq}{\begin{equation}}
\newcommand{\eeq}{\end{equation}}

\newcommand{\bc}{\begin{center}}
\newcommand{\ec}{\end{center}}
\newcommand{\bd}{\begin{displaymath}}
\newcommand{\ed}{\end{displaymath}}
\newcommand{\be}{\begin{equation}}
\newcommand{\ee}{\end{equation}}
\newcommand{\ba}{\begin{eqnarray}}
\newcommand{\ea}{\end{eqnarray}}
\newcommand{\bas}{\begin{eqnarray*}}
\newcommand{\eas}{\end{eqnarray*}}

\begin{document}

\title{Supermassive Stars in Quasar Disks}
\author{J. Goodman and Jonathan C. Tan}
\affil{Princeton University Observatory, Princeton, NJ 08544, USA.}



\begin{abstract}
We propose that supermassive stars may form in quasar accretion disks,
and we discuss possible observational consequences.  The structure and
stability of very massive stars are reviewed.  Because of high
accretion rates, quasar disks are massive and the fringes of their
optically luminous parts are prone to fragmentation.  Starting from a
few hundred solar masses, a dominant fragment will grow to the isolation
mass, which is a significant fraction of the disk mass, more quickly
than the fragment contracts onto the stellar main sequence.  A gap will
form in the disk and the star will migrate inward on the accretion
timescale, which is comparable to the star's main sequence lifetime.
By interrupting the gas supply to the inner disk, the gap may
temporarily dim and redden the quasar.  The final stages of stellar
migration will be a strong source of
low-frequency gravitational waves.
\end{abstract}


\section{Introduction}

Accretion disks massive
enough to fuel bright quasars are expected to be
self-gravitating beyond a few hundred to a few thousand
Schwarzschild radii ($\rs$), \citep[\emph{e.g. }][]{Shlosman_Begelman87}
Even under extreme assumptions,
it is difficult to see how such disks could extend beyond
$\sim0.1-1\pc$ without fragmenting completely into gravitationally
bound objects \citep{Goodman03, Sirko_Goodman03}.  Self-gravity is
less problematic beyond $\sim 10\pc$ because
the stellar bulge dominates the rotation curve, 
and because the disk eventually becomes optically thin.

Quasar accretion disks are not resolved and there are few direct
constraints on their structure.  The principal reason for believing
in them at all is that one does not know of any other plausible mode
of accretion that converts mass to radiation with high efficiency.
The overall size of the accretion flow is however constrained by
the need for sufficient surface area to radiate the observed continuum.
At wavelength $\lambda$, this area is
\[
\sim 10^{34}\left(\frac{\lambda L_\lambda}{10^{46}\ergs}\right)
\left(\frac{T_{B}}{2\times10^4\K}\right)^{-1}
\left(\frac{\lambda}{1{\rm \mu m}}\right)^3\cm^2,
\]
assuming incoherent emission by gas at temperature $T_B$.  Thus the
optically luminous part of the disk should extend to at least
$\sim 10^3\rs$, where self-gravity becomes important.

We shall assume that a conventional thin disk does exist and extends
to $10^3\rs$ with an accretion rate sufficient to support a typical
bright QSO of mass $\gtrsim 10^8\msun$ and luminosity $\gtrsim
10^{46}\ergs$.  We shall argue that a likely consequence of the
incipient self-gravity at the outer edge of such a disk is
formation of supermassive stars.  

The term ``supermassive'' is truly justified here.
While fragments may start
with only $\sim 10^2\msun$, they seem likely to grow to a significant
fraction of the disk mass, perhaps $\sim 10^5\msun$.  Yet it is quite
possible that even such extreme objects would escape notice until
appropriate gravitational-wave detectors are built (\S\ref{sec:observe}).

Supermassive stars of even larger masses ($\sim 10^8$) were once
proposed as models for quasars themselves \citep{Hoyle_Fowler63b,
Zeldovich_Novikov71}.  This notion has long been abandoned on grounds
of stability (\S\ref{sec:structure}), and more conclusively because 
statistical arguments demonstrate that quasars convert mass
to energy much more efficiently than nuclear fusion
\citep{Soltan82,Chokshi_Turner92,Yu_Tremaine02}.  It remains possible
that supermassive stars may be seeds or precursors to quasars
\citep{Rees84_ARAA}.

In this paper, we accept the conventional view that the energy source
for quasars and AGN is accretion onto massive black holes.  But the
disks required by this interpretation of the facts are so massive and
dynamically cool that they are likely to form substructures.  
We expect this to occur
\emph{via} the standard
local dynamical instability of self-gravitating disks, \emph{i.e.} the
gaseous version of the Toomre instability \citep{Toomre64}, although
growth of very large masses may perhaps be seeded by other means,
for example capture of stars (\S4.3).

In steady accretion, gravitational instability becomes progressively
more severe towards larger radii \citep[\emph{e.g.}][henceforth Paper
I]{Goodman03}.  Thus one might chose to consider a strongly unstable
region where the disk is likely to fragment completely: in other
words, where most of the mass would be found in clumps rather than a
smooth layer.  Others have investigated this regime
\citep{Shlosman_Begelman89,Shlosman_Begelman_Frank90, Kumar99}.
However, we have chosen to focus on a transitional region where
self-gravity is mild and bound structures, where they exist, are
likely to be well separated.  There are several reasons for our
choice.  For one thing, we doubt whether very strongly
self-gravitating material can form an accretion flow at all, except as
stars (Paper I).  
For another, there is little direct evidence that steady disk
accretion extends beyond $\sim 10^3\rs$, where self-gravity becomes
strong.  If it does, then a marginally self-gravitating region surely
exists by continuity.  Finally, marginal self-gravity offers the
methodological advantage that one can study it as a perturbation to
non-self-gravitating, thin-disk accretion, which is somewhat
understood.

The plan of the paper is as follows.  In \S\ref{sec:structure}, we
review the structure of supermassive stars and what is known about
their stability.  \S\ref{sec:initial} describes the physical
conditions of a QSO accretion disk in the region where it is likely to
be marginally self-gravitating, assuming an $\alpha$ disk in steady
state \citep{Shakura_Sunyaev73}.  Scalings with black hole mass,
accretion rate, and viscosity parameter are given.  \S4 builds upon
these standard results to discuss the formation, growth, and fate of
bound fragments (``stars'') that form in marginally self-gravitating
regions, with particular emphasis on their maximum mass.  \S5 sums up.

\section{Very massive stars}\label{sec:structure}

This section collects some elementary but important structural
formulae for very massive stars; somewhat more accurate forms were given by
\citet[][henceforth BAC]{Bond_Arnett_Carr84}.  We also summarize what is
known about the stability of such objects, both before and
after they reach the main sequence.

Very massive stars tend to be radiation-pressure dominated.  To the extent
that they are chemically homogeneous and the opacity ($\kappa$) is constant,
in hydrostatic and radiative (but not necessarily nuclear)
equilibrium, they are approximated by Eddington models:
\begin{eqnarray}\label{eq:Edmodel}
\pgas&=&\betas p~+~\mbox{constant},\nonumber\\
L_*&=& (1-\betas)L_{\rm Edd,*} = (1-\betas)\frac{4\pi GM_* c}{\kappa},
\end{eqnarray}
in which $\pgas$ and 
$\prad=p-\pgas$
are the partial pressures
of gas and radiation, and the asterisk is
used to distinguish parameters of the star from those of
the surrounding disk.
The dimensionless parameter $\betas<1$ is
nearly constant within a given star.  The additive constant on
the first line above is unimportant throughout most of the
interior, so that we may regard $\beta_*$ approximately as
$\pgas/p$.

Except near the surface where the
additive constant must be taken into account, and in the core
where luminosity may not be proportional to mass, it follows that
the stars are $n=3$ polytropes,
\begin{equation}\label{eq:polystate}
p\approx K\rho^{4/3},\quad K= \left[\frac{3}{a}
\left(\frac{\kB}{\mu\mH}\right)^{4}\,\frac{1-\betas}{\betas^4}\right]^{1/3},
\end{equation}
where $\mu$ is the molecular weight relative to hydrogen, so that
$\mu\approx 0.62$ for a fully ionized gas of solar metallicity.
From standard results for such polytropes,
\begin{equation}\label{eq:betaofM}
\frac{M_*}{\msun}\approx 47\,\frac{\sqrt{1-\betas}}{\betas^2}
\left(\frac{\mu_\odot}{\mu_*}\right)^2.
\end{equation}
It is important that this relation does not depend upon the stellar
radius ($R_*$), so that it holds even if the star has not reached
the main sequence.  Thus for extreme masses $M\gg 100\msun$, one has
$\betas\propto M^{-1/2}$.  The binding
energy
\begin{equation}\label{eq:BE}
-E_* = \frac{3}{4}\betas\frac{GM_*^2}{R_*}
\end{equation}
is small because the gravitational and internal energy nearly cancel,
and the Kelvin-Helmholtz time is correspondingly reduced:
\begin{equation}\label{eq:tKH}
\tkh\equiv\frac{|E_*|}{L_*}= \frac{3\kappa M_*}{16\pi c R_*}\,
\frac{\betas}{1-\betas}.
\end{equation}
The entropy per unit mass is
\be
\label{eq:entropy}
S=\frac{\kB}{2\mu\mH}\left[\frac{8(1-\betas)}{\betas}+
\ln\left(\frac{1-\betas}{\betas\mu^2\bar\rho}\right)\right]~+~\mbox{constant},
\ee
so that the effective adiabatic index for perturbations is
\begin{equation}\label{eq:adind}
\Gamma_1= \frac{32-24\betas-3\betas^2}{3(8-7\betas)}\approx\frac{4}{3}~+
\frac{1}{6}\betas~+O(\betas^2).  
\end{equation}

On the main sequence, the central temperature is nearly constant.
Since $GM_*^2/R_*^4\propto p_{\rm c}\propto T_{\rm c}^4/(1-\betas)$,
one has $R_*\propto M_*^{1/2}$.  More accurately (BAC)
\begin{equation}\label{eq:RofM}
\frac{R_{*,\rm ms}}{R_\odot}= 10\,\frac{(1-\betas)^{0.39}}{\betas^{0.95}}
\left(\frac{X_{CN}}{0.01}\right)^{0.05} ~\rightarrow 
~1.6\left(\frac{M_*}{\msun}\right)^{0.47}
\left(\frac{\mu_*}{\mu_\odot}\right)^{0.95} (M_*\gg 10^3\msun),
\end{equation}
where $X_{CN}$ is the mass fraction in CNO elements.
The Kelvin-Helmholtz time on the main sequence is therefore
\begin{equation}\label{eq:tkhms}
t_{\rm KH,ms}\approx 3300\,\left(\frac{\kappa}{0.4\g\cm^{-2}}\right)
\left(\frac{\mu_\odot}{\mu_*}\right)^2
\betas^{-0.053}(1-\betas)^{-0.89}\yr.
\end{equation}
The dependence on $M_*$, which enters through $\betas$ and eq. (\ref{eq:betaofM}),
is so weak that we shall generally ignore it.

The main-sequence lifetime is also nearly constant (BAC),
\begin{equation}\label{eq:tms}
t_{\rm ms}\approx (2-3)\,\times 10^6\yr~\sim\frac{\veps_{\rm fusion}
M_*c^2}{L_{\rm Edd,*}}~,
\end{equation}
unless cut short by catastrophic instabilities.

\subsection{Stability of massive stars}\label{sec:stability}

Instabilities can be categorized according to if they occur before, during,
or after the main-sequence phase.

The most famous pre-main-sequence instability is a
relativistic one.  Since the adiabatic index (eq. \ref{eq:adind}) is
close to $4/3$, changes in internal and gravitational energy are
almost equal and opposite if the star is homologously and adiabatically
compressed.  Small corrections to the energy can have a large effect
on stability.  One such correction is general relativity, which
increases (in absolute magnitude) the potential energy
by a fraction $\sim (GM_*/c^2 R_*)$.
A very crude estimate of the minimum unstable mass
can be obtained by evaluating $GM_*/R_*$ on the main sequence and
equating this to $\beta_*/6$.  Using equations (\ref{eq:RofM}) and (\ref{eq:betaofM}),
this yields $M_{*,GR}\sim 6\times 10^5\msun$.  More detailed calculations
predict a rather smaller mass \citep{Chandra64},
$M_{*,GR}\approx 5\times 10^4\msun$, for nonrotating stars.

Rotation lends
stability because its energy scales with $\pgas$
rather than $\prad$ under homologous contractions, that is, 
$E_{\rm rot}\propto R_*^{-2}\propto\rho_*^{2/3}$.  The maximum
rotational energy of a uniformly rotating $n=3$ polytrope is small,
$E_{\rm rot}/|E|\lesssim 0.007$, at which value centrifugal and
gravitational forces balance at the equator
\citep{Zeldovich_Novikov71}.
General-relativistic effects are also small, however, so  rotation
is important, especially since stars forming
out of a disk will probably rotate strongly.
We therefore take the minimum mass for general-relativistic instability
to be the value appropriate to maximal uniform rotation
\citep{Baumgarte_Shapiro99}
\begin{equation}\label{eq:MGR}
M_{*,GR}\approx 10^6\msun.
\end{equation}

Stars with $M_*<M_{*,GR}$ reach the main sequence on a timescale no
longer than the Kelvin time (eq. \ref{eq:tkhms}).  Thermonuclear reactions commence on the
CNO cycle, since quasar accretion disks are at least as metal rich as the
Sun \citep[and references therein]{Dietrich03}.
At $M_*\gtrsim 100\msun$, the sensitivity of the
thermonuclear reaction rate to central temperature drives a linear
overstability to pulsations
\citep{Eddington26,Ledoux41,Schwarzschild_Harm59}.  Non-linear
calculations suggest that this instability drives mass loss
\citep{Appenzeller70a,Appenzeller70b, Papaloizou73}. The rate of
mass loss is uncertain, but is probably no greater than
$\dot{m}_*=5\times 10^{-5}\msun\,\yr^{-1}$ for a $130\msun$ star and
$\dot{m}_*=5\times 10^{-4}\msun\,\yr^{-1}$ for a $270\msun$ star
\citep{Appenzeller70a}, and could be as little as $\sim
10^{-6}\msun\,\yr^{-1}$ for similar masses \citep{Papaloizou73}.
The scaling of $\dot m_*$ to much larger stellar masses is also
uncertain.

The above mass-loss rates may be compared to those resulting from
radiation pressure driven winds. Consider a $120\msun$ zero age main
sequence (ZAMS) star with $T_{\rm eff}=5.3\times 10^4\:{\rm K}$ and
luminosity $L_*=1.8\times 10^6\:L_\odot$ \citep{Schaller_etal92}.  The
analytic relation of \citet{Vink_etal00} for radiation pressure driven
winds including multiple scattering, predicts a mass-loss rate of
$1.65\times 10^{-5}\msun\,\yr^{-1}$. This mass-loss rate scales
approximately as $L_*^{2.2}$ and $m_*^{-1.3}$, with an additional
weaker increase resulting from hotter surface temperatures. 
It can be considered as a lower limit to the actual mass-loss rates
from massive stars, since these typically
increase during the later periods of stellar evolution, particularly
during the Wolf-Rayet phase. For example, at the above rate the stellar mass
would decrease to about $70\msun$ in the lifetime of $\sim 3\:{\rm
Myr}$. However, more detailed evolutionary models including mass loss
rates calculated consistently for each stage of the evolution give a
final stellar mass of only $\sim 8\msun$ \citep{Schaller_etal92}; half
of the total mass loss occurs in the last 0.5~Myr of evolution. It
should be noted that these results are for non-rotating stars,
evolving with no additional mass accretion.

If the star manages to survive the main-sequence, then in the
post-main-sequence phase
an instability arises by the creation of $e^+e^-$ pairs when
the central temperature $\sim 10^8\K$ \citep{Zeldovich_Novikov71}.
This is believed to result in an explosion if $M_* \lesssim 200\msun$
(BAC) but complete collapse to a black hole if $M_* \gtrsim 300\msun$
\citep{Fryer_Woosley_Heger01}.

\section{Initial conditions for massive star formation}\label{sec:initial}

Several physical properties differentiate the star-forming
environment of a quasar disk from that of a typical region of Galactic
massive star formation \citep{Plume_etal97,McKee_Tan03}. 
Firstly the densities and pressures are many
orders of magnitude greater. Secondly the environment is subject to
strong shear, due to the Keplerian orbits around the massive black
hole. 
The dominant source of heating of the gas, at least in the inner regions 
of the disk, is due to its accretion in the QSO's potential.
At the typical temperatures in the QSO disk we
expect dust grains and molecules to have been destroyed, and finally, radiation
pressure can lend significant support to the gas.

\subsection{Physical conditions in the disk}\label{subsec:physcon}

From Paper I, we have the following properties for geometrically
thin, optically thick, Keplerian disks that are heated by viscous dissipation
only. The radial dependence of the surface density, $\Sigma$, and the
midplane temperature, $T$, are given by:
\ba\label{Tmid}
T&=& \left(\frac{\kappa \mu m_H}{16\upi^2\alpha\beta^{b-1}\kB\sigma}\right)^{1/5}
\dot M^{2/5}\Omega^{3/5}\\
&\approx& 5.27\times10^4\left(\frac{\lum^2\krel\mu}{\veps_{0.1}^2\alpha_{0.3}
\beta^{b-1}}\right)^{1/5} M_8^{-1/5}r_3^{-9/10}
\,\K,\nonumber
\ea
\ba\label{Sigmar}
\Sigma&=& \frac{2^{4/5}}{3\upi^{3/5}}
\left(\frac{\mu^4 m_H^4\sigma}{\kB^4}\right)^{1/5}
(\alpha\beta^{{b}-1})^{-4/5} \kappa^{-1/5}\dot M^{3/5}\Omega^{2/5}\\
&\approx& 2.56\times 10^5\,(\alpha_{0.3}\beta^{b-1})^{-4/5}
\lum^{3/5}\veps_{0.1}^{-3/5}\krel^{-1/5} \mu^{4/5} M_8^{1/5}
r_3^{-3/5}~\mbox{g cm}^{-2}\,,
\nonumber
\ea
The meaning of the various symbols is as follows: $\Omega=(GM/r^3)^{-1/2}$,
$\krel$ is the opacity relative to the electron-scattering value
($0.4\cm^2\g^{-1}$);
$\mu$ is the mean particle mass in units of the Hydrogen atom mass, $m_H$;
$0.3\apn$ is the Shakura-Sunyaev viscosity parameter [scaled to the
largest likely value in a selfgravitating, but not fragmented, thin
disk \citep{Gammie01}];
$\beta\equiv p_{\rm gas}/p$;
$0.1\epn$ is the radiative efficiency of the disk ($=L/\dot Mc^2$);
$\lum$ is the disk luminosity relative to the Eddington limit;
$M=10^8 M_8\msun$ is the black hole mass; 
and $\rs=2GM/c^2\approx 10^{-5} M_8\pc$ is the Schwarzschild radius,
with $r_3$ the orbital radius in units of $10^3\rs$.
Also $b$ takes a value of either 0 or 1, depending on whether the 
viscosity is proportional to the total or gas pressure, respectively. 

By including the opacity as an explicit parameter, we avoid its
sometimes complicated functional dependence on density and temperature,
thereby obtaining simple algebraic formulae for the disk properties.
The numerical disk models of \citet[henceforth Paper II]{Sirko_Goodman03},
which took realistic opacities, show that $\kappa\sim 1\unit{g\,cm^{-3}}$
throughout the region of the disk of interest to us, namely
$r\lesssim 10^3\rs$.  (Indeed, $\hat\kappa\sim 1$  out to $r\sim 10^4\rs$
if nonviscous auxiliary heating maintains $Q\approx 1$ to that radius.)
Furthermore, as will be seen, many quantities of interest depend rather
weakly on $\hat\kappa$.

If viscosity scales with gas pressure (${b}=1$) then 
eqs.~(\ref{Tmid}) \& (\ref{Sigmar}) do not depend on $\beta$, which in
any case is a known function of density and temperature:
\bd
\frac{\beta}{1-\beta}=\frac{p_{\rm gas}}{p_{\rm rad}}=
\frac{3c\kB}{4\sigma m}\,\frac{\rho}{T^3}= \frac{3c}{8\sigma}
\left(\frac{\kB}{m}\right)^{1/2}\beta^{1/2}\frac{\Sigma\Omega}{T^{7/2}}\,;
\ed
using eqs.~(\ref{Tmid}) \& (\ref{Sigmar}),
\begin{eqnarray}\label{eq:quartic}
\frac{\beta^{(1/2)+({b}-1)/10}}{1-\beta}&=&
(2^3\upi^4)^{1/5}\alpha^{-1/10}
c(\kB/m)^{2/5}\sigma^{-1/10}
\kappa^{-9/10}\Omega^{-7/10}\dot M^{-4/5}\nonumber\\[5pt]
&\approx&
0.311\,\apn^{-1/10}\krel^{-9/10}
\mu^{-2/5}(\epn/\lum)^{4/5} M_8^{-1/10}\,r_3^{21/20}.
\end{eqnarray}
Using eq.~(\ref{Tmid}) to eliminate the isothermal sound speed,
$c_s=\sqrt{p/\rho}=kT/(\mu m_H)$ from \citet{Toomre64}'s gravitational
stability parameter yields
\ba\label{Qgpd}
Q&=& 3(4\upi)^{-3/5}\alpha^{7/10}\beta^{(7{b}-12)/10}
\left(\frac{\kB}{\mu m_H}\right)^{6/5}\sigma^{-3/10}G^{-1}
\kappa^{3/10}\dot M^{-2/5}\Omega^{9/10}\nonumber\\
&\approx& 1.0\,\alpha_{0.3}^{7/10}\beta^{(7{b}-12)/10}
\veps_{0.1}^{2/5}\lum^{-2/5}\krel^{3/10} \mu^{-6/5}
M_8^{-13/10}r_3^{27/20}~.
\ea

Consider quasar accretion disks at the
inner radius of gravitational instability where $Q=1$:
\be
\label{eq:rsg}
\rsg \approx 915\,\apn^{14/27}\krel^{2/9}\mu^{-8/9}(\epn/\lum)^{8/27}
M_8^{-26/27}\beta^{(14b-24)/27}\rs.
\ee
This equation differs slightly from
eq.~(10) of \citet{Goodman03} because some approximations assuming
$\beta\ll 1$ were used there.

Substituting for $r$ from eq. (\ref{eq:rsg}) into eq. (\ref{eq:quartic}) yields
\be\label{eq:betasg}
\frac{\beta_{\rm s.g.}^{1\,-b/3}}{(1-\beta_{\rm s.g.})^{3/4}}=0.389\,
\apn^{1/3}\krel^{-1/2}\mu^{-1}(\epn/\lum)^{5/6} M_8^{-5/6}\,.
\ee
Thus gas and radiation pressure are comparable at $r_{\rm s.g.}$.
Equations~(\ref{eq:betasg}) \& (\ref{eq:rsg}) imply $r_{\rm s.g.}$ is
not very sensitive to the choice of viscosity law, \emph{e.g.}
$r_{\rm s.g}\approx 2700\rs$ for $b=0$ \emph{vs.}
$1700\rs$ for $b=1$ at the fiducial values of the other parameters.
On the other hand,  radiation pressure is at least moderately
dominant at all $r\le r_{\rm s.g.}$, so that we may replace
$1-\beta$ with unity in the quartic (\ref{eq:quartic}); this
makes $\beta$ a simple power law in all other parameters.
To further simplify the discussion, we shall also take the
viscosity to be proportional to gas pressure except
where stated otherwise.

With these assumptions ($\beta\ll 1$, $b=1$), the properties of the
disk at $r\le r_{\rm s.g.}$ are as follows.  The midplane density is
\be\label{eq:rhovsr}
\rho=\frac{\Sigma\Omega}{2\cs} \approx 4.31\times 10^{-10}\,
\apn^{-4/5}\krel^{-6/5}\mu^{4/5}(\epn/\lum)^{2/5}M_8^{-4/5} r_3^{-3/5}
\g\cm^{-3}\,
\ee
(corresponding to number densities of hydrogen nuclei
$\sim 10^{14}\:{\rm cm^{-3}}$), and the total pressure is
\be\label{eq:pvsr}
p/\kB \approx 1.41\times 10^{20}\,
\apn^{4/5}\krel^{4/5}\mu^{-4/5}(\lum/\epn)^{8/5}M_8^{4/5} r_3^{-18/5}
\cm^{-3}\K\,.
\ee
The midplane temperature is given by eq.~(\ref{Tmid}) with $b=1$
(note that at this temperature dust and molecules will be destroyed),
which corresponds to an isothermal sound speed, $c_s=\sqrt{p/\rho}$, of
\be\label{eq:csvsr}
\cs \approx 67.2\,
\krel^{1/5}\lum\epn^{-1}r_3^{-9/10}
\kms\,.
\ee
The disk scale height is
\be\label{eq:hvsr}
h =\frac{\cs}{\Omega} \approx 10\,\krel\lum\epn^{-1}\rs\,,
\ee
the surface density is given by eq.~(\ref{Sigmar}),
and the total disk mass inside $r$ is
\be
\label{eq:Mdisk}
M_{\rm disk}(<r) \approx 5.03\times 10^{5}\,
\apn^{-4/5}\krel^{-1/5}\mu^{4/5}(\lum/\epn)^{3/5}M_8^{11/5} r_3^{7/5}
\msun\,.
\ee
The orbital timescale is
\be\label{eq:torb}
t_{\rm orb} \approx 8.79\, M_8 r_3^{3/2}\yr\,,
\ee
whereas the accretion time is
\be\label{eq:tvisc}
t_{\rm visc}=\frac{M_{\rm disk}(<r)}{\dot M}\approx
2.28\times 10^5\,
\apn^{-4/5}\krel^{-1/5}\mu^{4/5}(\epn/\lum)^{2/5}M_8^{6/5} r_3^{7/5}
\yr\,,
\ee
and the circular velocity is
\be\label{eq:vcirc}
v_{\rm circ}\approx 6710\,r_3^{-1/2}\kms.
\ee
Finally, we may eliminate $\beta$ from eq.~(\ref{eq:rsg}) to express the
radius of marginal self-gravity itself as
\be\label{eq:rsg_smallbeta}
\rsg\approx 1550 (\apn/\mu)^{1/3}(\lum/\epn)^{1/6}\krel^{1/2}M_8^{-1/2}\,\rs.
\ee

It is worth emphasizing once again that eqs.~(\ref{eq:rhovsr})-(\ref{eq:rsg_smallbeta})
assume $\nu\propto\pgas$ and $\beta\ll1$ at $r\le\rsg$; in view of 
eq.~(\ref{eq:betasg}), the latter condition is expected to be
satisfied at high accretion rates, $\dot M\gtrsim 1\,M_\odot\,\yr^{-1}$.

\subsection{Fragmentation}

We assume that the initial sizes and masses of protostars forming
in the disk are determined by \citet{Toomre64}'s dynamical
instability.
The shear stabilizes long wavelength perturbations from collapse,
while the pressure stabilizes short wavelength modes.
The most unstable mode has radial wave number $k_{\rm m.u.}=\pi G
\Sigma/c_s^2 = (Qh)^{-1}$ so that
\be\label{eq:lmu}
\lambda_{\rm m.u.}(r_{\rm s.g.}) =
\frac{2 \pi}{k_{\rm m.u.}} \approx 63\,\krel(\lum/\epn)\rs.
\ee
The numerical expression assumes $\beta_{\rm s.g.}\ll 1$ so that
$h$ is given by eq. (\ref{eq:hvsr}), which is probably a good approximation
for $M_8\gtrsim 1$ and $\lum\sim 1$.
The instability is axisymmetric, but
fragments having comparable
azimuthal and radial dimensions probably result.
The corresponding mass at $r=\rsg$ is then of order
\be\label{eq:Mmu}
M_{\rm m.u.} \equiv \Sigma \lambda_{\rm m.u.}^2 \sim
300\,\apn^{-1}\krel^{3/2}\mu(\lum M_8/\epn)^{5/2}\msun.
\ee

A second way to estimate the fragment mass is to note that
$\beta$ is approximately conserved under
an adiabatic contraction, since as eq.~(\ref{eq:entropy}) shows,
the entropy depends only logarithmically on density at fixed $\beta$.
Therefore one may set
$\betas\to\beta(r_{\rm s.g.})$ in the relation eq. (\ref{eq:betaofM}) for
Eddington stellar models.
This yields an initial stellar mass about four times larger than
eq. (\ref{eq:Mmu}) but with the same scalings.
The star is likely to gain a great deal more mass by accretion,
as we shall explain next.

\section{Ultimate masses}\label{sec:ultimate}

In a strongly unstable case where $Q<1$, the disk might break up
completely into closely spaced fragments.  The subsequent evolution
would then be dominated by collisional agglomeration.  Let us consider
instead a less unstable situation ($Q\gtrsim 1$) so that the initial
fragments are well separated and contain only a small fraction of the
disk mass.  Then a protostar gains mass mainly by accretion from
the ambient gaseous disk.

We first consider whether the fragment (with mass given by eq. \ref{eq:Mmu}) 
will itself
fragment into smaller masses.  The usual condition for this to happen
is that the cooling time be less than the the dynamical time.  The
initial density ($\rho_*$) and optical depth ($\tau_*\sim \kappa\rho_*
R_*$) of the fragment will be comparable to those of the ambient
disk, and its radius $R_*\sim\lambda_{\rm m.u.}\sim 2h$, the disk
half-thickness.  The initial dynamical time is
$(G\rho_*)^{-1/2}\sim (G\rho_{\rm disk})^{-1/2}=Q \Omega$, and the
cooling time of the fragment is of order the local thermal time of
the disk, $t_{\rm th}=\alpha^{-1}\Omega$ \citep{Pringle81}.
Therefore, provided $\alpha Q<1$ then $t_{\rm cool}> t_{\rm dyn}$.
Furthermore, if the fragment contracts at constant mass, then $t_{\rm
dyn}\propto R_*^{3/2}$, whereas $t_{\rm cool}\propto R_*^{-1}$ if
$\kappa$ is constant, so that $t_{\rm cool}/t_{\rm dyn}\propto R_*^{-5/2}$,
making further fragmentation increasingly unlikely as the radius shrinks.
Even if subfragmentation were to begin, rapid accretion from the
surrounding disk (see below) would probably drive the fragments
together again by absorbing energy and angular momentum from the
relative motion of the fragments.  In short, the mass of the protostar
is more likely to grow by accretion than to decrease by fragmentation.
Nevertheless, it would be useful to confirm this by numerical
simulations in two or three dimensions with allowance for cooling.

\subsection{Accretion and the isolation mass}

The accretion rate will be estimated by analogy with growth of
terrestrial planets in a planetesimal disk \citep{Lissauer87}.
An important lengthscale in the latter problem is the Hill radius
$\rh\equiv (M_{\rm p}/3M_*)^{1/3}r$,
which is essentially the
size of the Roche lobe surrounding the planet (mass $M_{\rm p}$) at
an orbital distance $r$ from the central star (mass $M_*$).
At distances $\ll\rh$ from the planet, motions of planetesimals
are dominated by the gravitational field of the planet rather than
the star. In a quasar disk, the role of the planet is played by the protostar 
(mass $M_*\sim 10^{2-5}\msun$),
and that of the Sun, by the central black hole (mass $M\gtrsim 10^8\msun$).
Thus
\be\label{eq:RH}
\rh \equiv \left(\frac{M_*}{3M}\right)^{1/3} r.
\ee

\begin{figure}
\epsscale{0.8}
\plotone{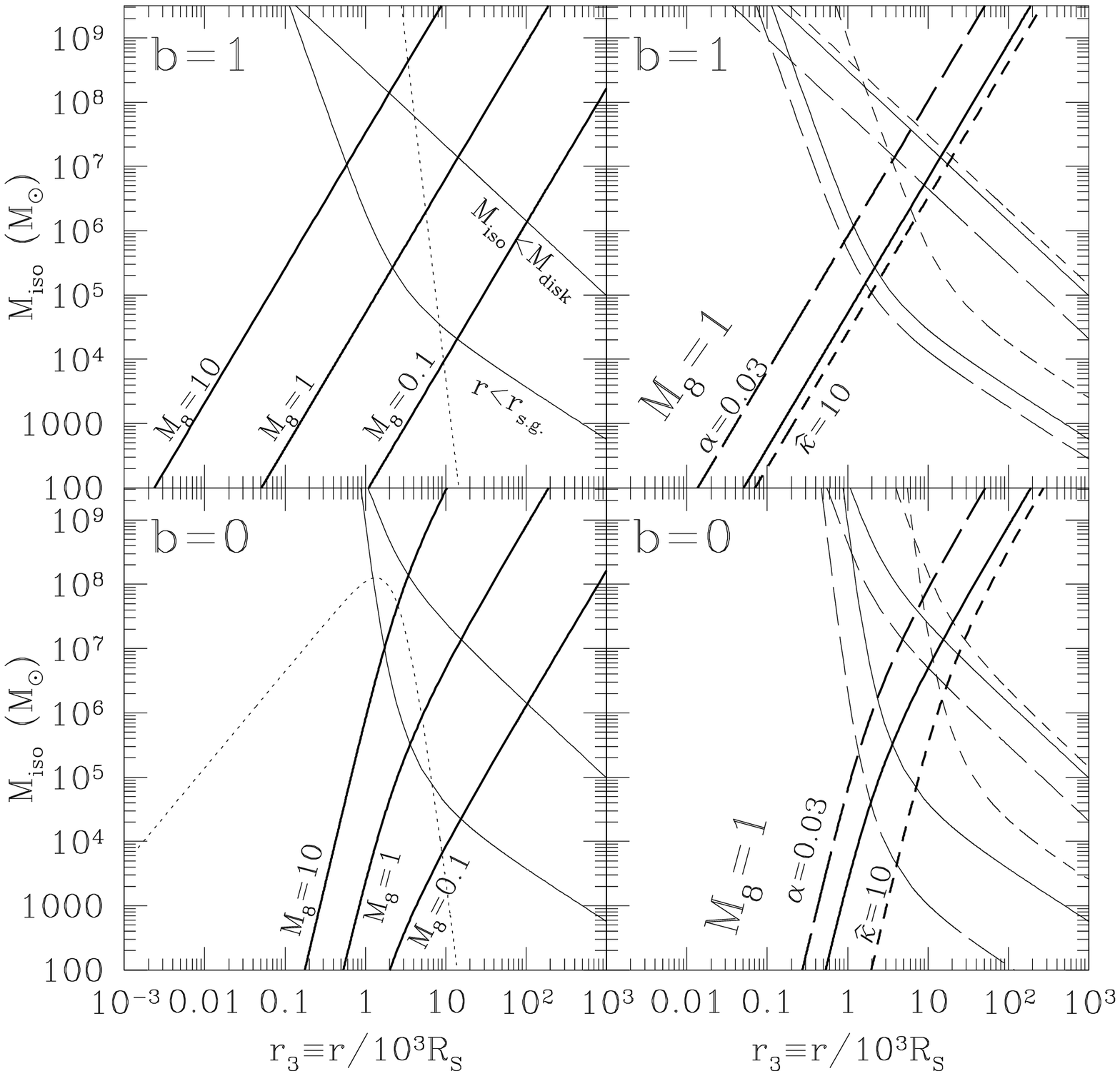}
\caption{
\label{fig:miso}
\footnotesize Isolation mass, $M_{\rm iso}$ (see eq. \ref{eq:Miso}),
versus radial location (in Schwarzschild radii) in the disk (heavy
lines) with $b=0,1$ (bottom, top panels) for $10^7, 10^8, 10^9\sm$
central black holes with $\krel=1$, $\alpha=0.3$, $\mu=0.6$, $\lum =
\epn = 1$, and $f_H=1$ (left panels). In each panel the lower thin
line shows the locus defined by $r=r_{\rm s.g.}$ (eq. \ref{eq:rsg}),
with each point on the line representing a different central black
hole mass. Similarly, the upper thin line shows the condition when
$M_{\rm iso}(r)=M_{\rm disk}(r)$ (here we define $M_{\rm
  disk}(r)\equiv (10/7)\pi \Sigma^2$ as a measure of the maximum mass
available for star formation from the disk). The dotted line traces
the conditions where the midplane temperature is $10^4\:{\rm K}$ ---
in the cooler region we may expect significant departures from the
fiducial opacity.  The right panels show the effects on $M_{\rm
  iso}(r)$ for the $M_8=1$ case of increasing $\krel$ by a factor of
10 (heavy dashed line) and reducing $\alpha$ by a factor of 10 (heavy
long dashed line). The effects on the self-gravity radius condition
and the disk mass condition are also shown with the same line types.
Note that $M_{\rm iso}\propto f_H^{3/2}$, which introduces additional
uncertainty.  We expect supermassive star formation to the isolation
mass at $r\simeq r_{\rm s.g.}$, and from these figures we can see that
this is not prevented by a lack of material from the disk for the
range of parameters we have considered.
}
\end{figure}

\begin{figure}
\epsscale{1.0}
\plotone{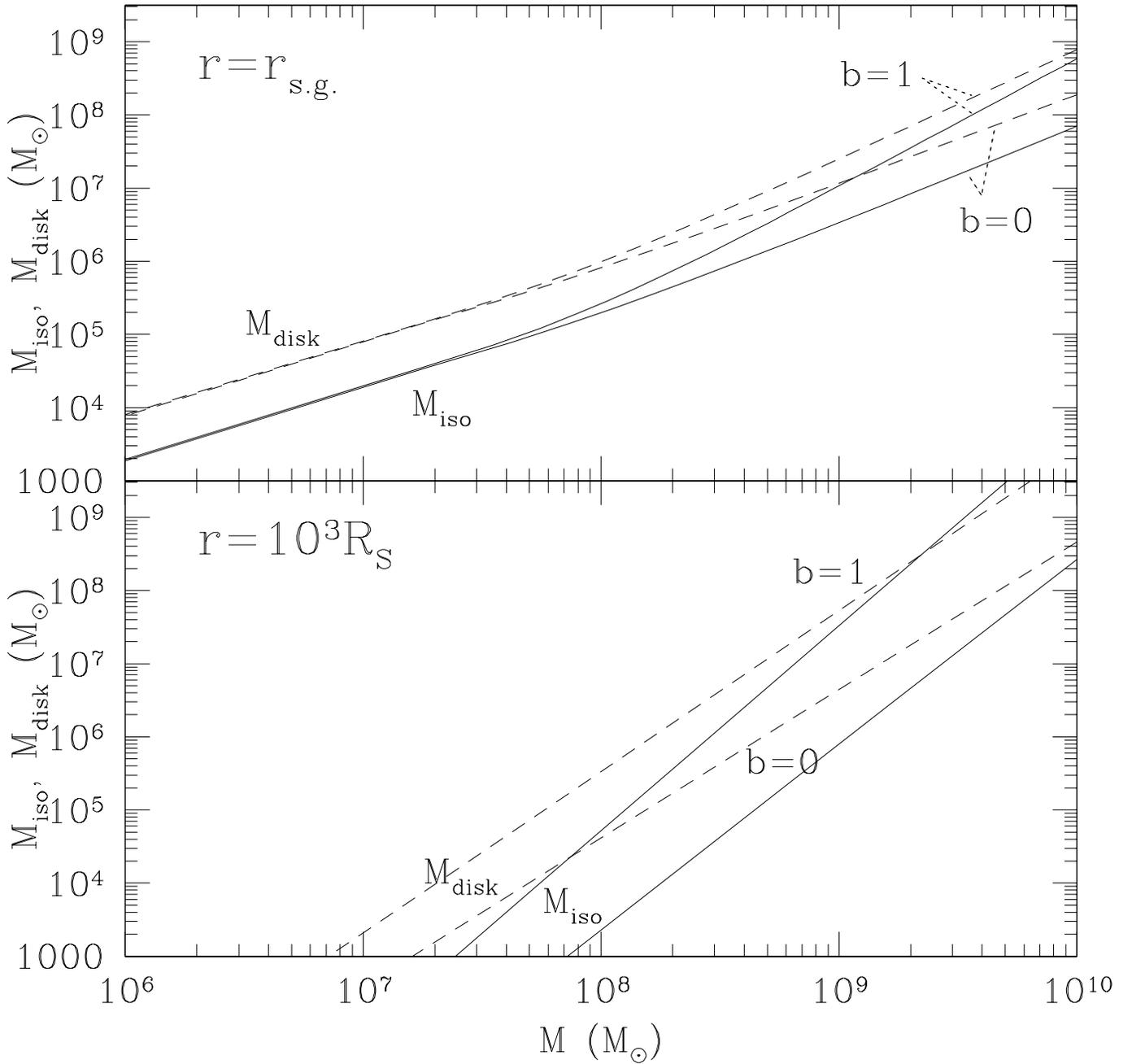}
\caption{
\label{fig:miso2}
Isolation mass at $r=r_{\rm s.g}$ (top panel) and $10^3 R_S$ (bottom)
versus central black hole mass, assuming $\alpha=0.3$, $\krel=1$,
$\lum = \epn = 1$, and $f_H=1$.  The cases with $b=0,1$ are shown, as
labeled. The dashed lines show the disk mass inside $r$.
}
\end{figure}

Tidal torques exerted by the growing protostar will eventually
open a gap in the disk, shutting off or at least reducing the rate
of accretion.  These torques, however, act only on gas
that has encountered the protostar at least once.  Material on an
orbit whose semimajor axis differs from that of the protostar by
$\sim\rh$ stands a good chance of being accreted on its first passage.
Let us therefore assume that the protostar
accretes the entire annulus $|r-r_*|\le f_H\rh$, where $f_H\sim O(1)$.
(Lissauer infers $f_H\sim 3-4$ from simulations of planetesimal growth.)
The mass of this annulus is larger than the initial mass
(eq. \ref{eq:Mmu}) of the protostar by a large factor $\sim r/h$ if $Q\sim 1$.
Therefore, the initial width of the gap that is cleared is proportional to
Hill radius of the mass accreted.
This condition defines the {\it isolation mass} \citep{Lissauer87},
\be\label{eq:Miso}
M_{\rm iso} = \frac{(4\pi r^2\Sigma f_H)^{3/2}}{(3M)^{1/2}}~
\approx 0.96\times 10^5\,f_H^{3/2}\alpha_{0.3}^{-6/5} \beta^{6(1-b)/5} \krel^{-3/10}
\mu^{6/5}(\lum/\epn)^{9/10}M_8^{14/5}r_3^{21/10}\msun.
\ee
Numerically, if $f_H\gtrsim 1$,
this is comparable to the disk mass (eq. \ref{eq:Mdisk}),
but formally $M_{\rm iso}\propto M_{\rm disk}\times (M_{\rm disk}/M)^{1/2}$.
The dependence of
$M_{\rm iso}$ with $r$ and $M$ is shown in Figures
\ref{fig:miso} and \ref{fig:miso2}.
The isolation mass is startlingly large and one wants
to examine very critically the assumptions that have led to it.

The dynamics of a pressure-supported
gas are somewhat different from those of a nearly collisionless swarm of
planetesimals.  \citet{Rafikov03c} has recently demonstrated that the
growth rate of a planetary embryo slows significantly at a mass
$\ll M_{\rm iso}$
because gravitational encounters with the embryo tend to ``heat'' the
epicyclic velocity dispersion of the planetesimals and thereby decrease
their rate of collisions with the embryo.  In this respect a quasar disk
is more favorable for rapid growth because the gas can cool radiatively.
In fact, the cooling time is $\sim (\alpha\Omega)^{-1}$, which is
probably short compared to the time between successive encounters
with the protostar, $\sim 2\pi(\Omega f_H\rh/r)^{-1}$.
Also, the accreting protostar probably nearly fills its
Hill sphere during the growth phase (see below) so that it presents a
relatively large cross section for ``collisions'' compared to a solid
planetary embryo whose density is determined by chemical bonds.

On the other hand, whereas impacts of planetesimals on planetary
embryos are perhaps fully inelastic, it is not immediately
obvious that the accreting protostar can accept gas at the
rate that it arrives \emph{via} differential rotation.
To address this question, we compare the rate at which mass
enters the Hill sphere, the rate at which this material
sheds its ``spin'' angular momentum, and the rate at
which it sheds energy.  Since masses much larger than (\ref{eq:Mmu})
are concerned, we estimate these rates in the limit $\rh\gg H$.

We begin by estimating the rate of growth of the Hill sphere.
Material on an orbit at radial separation $\Delta r$
from the center of the Hill sphere approaches it at azimuthal speed
\[
v_{rel} =  r\frac{d\Omega}{ dr}\Delta r\approx 
-\frac{3}{ 2}\Omega\Delta r.
\]
The surface density on such a streamline is $\Sigma$, the background
surface density of the disk, if the material is
approaching the star for the first time, that is to say,
\be\label{eq:trH}
t < \frac{2\pi r }{ |v_{rel}|} = \frac{4\pi}{3}\frac{r}{ f_H \rh}\Omega^{-1},
\ee
where $t$ is the time since accretion begins.  We have set $\Delta r=
f_H\rh$ because if the inequality is satisfied at this separation,
then it is satisfied at smaller separations, and more distant material
is not likely to be much perturbed.
Hence mass enters the Hill sphere at the rate
\[
\frac{dM_H}{dt}\approx \Sigma\int\limits_{r-f_H\rh}^{r+f_H\rh} |v_{rel}|dr
\approx \frac{3}{2}\Sigma\Omega f_H^2\rh^2.
\]
Since $\rh\equiv r(M_H/3M)^{1/3}$, this can be rewritten as
\be\label{eq:rHdot}
\frac{d}{dt}\rh= \frac{\Omega\Sigma r^3}{6M}~=~{\rm constant.}
\ee
(We use $M_H$ rather than $M_*$ to denote the mass within the
Hill sphere so as not to prejudge the question as to the timescale
on which this mass concentrates into a quasi-spherical star rather than
a centrifugally supported disk: see below.)
Thus $\rh$ increases linearly with time until the two sides
of the inequality (\ref{eq:trH}) become equal,
at which point accretion presumably slows or stops.   This defines
the asymptote of $\rh$  and hence the isolation mass:
\begin{eqnarray*}
\rh(M_{\rm iso})&=& \left(\frac{2\pi\Sigma r^2}{9M}\right)^{1/2}r\,,\\
M_{\rm iso} &=& \frac{(2\pi f_H\Sigma r^2)^{3/2}}{9M^{1/2}}\,.
\end{eqnarray*}
The mass is smaller than our previous estimate (\ref{eq:Miso})
by a factor $6^{-3/2}$, which is an indication of the crudeness of both
arguments.  The approximate time at which this mass is achieved is found
by interpreting eq.~(\ref{eq:trH}) as an equality with $\rh$ as given above:
\begin{eqnarray}\label{eq:tiso}
t_{\rm iso} &\approx& \frac{2\pi}{\Omega}\left(\frac{2M}{f_H^3\pi\Sigma r^2}
\right)^{1/2}\nonumber\\
&\approx& 210\,
f_H^{-3/2}\apn^{2/5}\krel^{1/10}\mu^{-2/5}(\epn/\lum)^{3/10}M_8^{2/5}
r_3^{4/5}\yr\,.
\end{eqnarray}
Since $M/\Sigma r^2\sim r/H$ at $Q=1$, $t_{\rm iso}$
is longer than the local orbital period by $\sim (r/H)^{1/2}$.

Having estimated the timescale on which the isolation mass accumulates
within its Hill sphere, we now consider the timescale on which it
contracts to radii $\ll\rh$.  Unless the contraction time is smaller
than the accumulation time, it is questionable whether the Hill sphere
can accept the full mass $M_{\rm iso}$ in the first place.  On the
other hand, a contraction time much shorter than $t_{\rm iso}$ would
suggest fragmentation.

In order to contract, the mass $M_H$ must shed both energy and
the angular momentum associated with rotation around its own axis
\emph{i.e.} spin.  We will now show that the timescale for
loss of angular momentum is likely to be shorter than that for
loss of energy.  Suppose the contrary: then the material within
the Hill sphere settles into a centrifugally supported disk,
which will be called the ``stellar disk'' to distinguish it from
the larger disk that orbits the black hole.  Let the mass-weighted
average thickness of this disk be $\bar h$, presumed to be $\ll\rh$
but not necessarily equal to $H$, the thickness of the ambient
black-hole disk;
let its median (half-mass) radius be $\bar R$; and let its median
angular velocity around its own axis be $\bar\omega$.
Then the viscous time in the stellar disk is
$(\alpha\bar\omega)^{-1}(\bar R/\bar h)^2$.
Suppose provisionally that this is long compared to $t_{\rm iso}$, as it
would be if $\bar R\sim\rh$ and $\bar h\lesssim H$.
Then $\bar R\propto\rh$ and $\bar\omega\propto\Omega$ as $\rh$ expands, since
the specific angular momentum of the material falling into the Hill
sphere is $\sim \Omega\rh^2$.
[In fact, the stellar disk
should be no larger than the streamline-crossing radius, which
is $R_{\rm sc}\approx 0.4\rh$ at these extreme mass ratios
$M_H:M\lesssim 10^{-3}$ \citep{Paczynski77}.  At and beyond $R_{\rm sc}$,
angular momentum is efficiently removed by the tidal potential.]
Thus $\bar h\propto\bar c_s/\Omega$, where $\bar c_s$ is the median
sound speed in the stellar disk.
The cooling rate of the disk by vertical radiative diffusion is therefore
\be\label{eq:tcooldisk}
\bar t_{\rm cool}^{-1}= f_1\frac{c}{\kappa\bar\Sigma \bar h}\,,
\ee
where $f_1$ is a positive dimensionless constant, $c$ the speed of light,
$\kappa$ the opacity,  and $\bar\Sigma$ the median
surface density.  We have assumed that the stellar disk is
radiation-pressure dominated, like the ambient black-hole disk, but
if not, then the cooling time would be
longer by a factor $(1-\bar\beta)^{-1}$.

New material joining the disk does so through a
shock that endows it with internal energy $\propto GM_H/\bar R$ per
unit mass.  Thus the median internal energy per unit area of the
circumstellar disk evolves as
\be\label{eq:fatdisk}
\frac{d}{dt}\bar\Sigma\bar c_s^2 = 
- f_1\frac{c}{\kappa \bar h}\bar c_s^2~+~
f_2\frac{GM_H\dot M_H}{\bar R^3}\,,
\ee
in which $f_2$ is another positive constant.  From the discussion
above, $\bar R\propto\rh\propto t$,
$M_H\propto t^3$, and therefore $\dot M_H\propto t^2$ and
$\bar\Sigma\propto M_H/\rh^2\propto t$.  So if
$\bar h$ were constant in time, then $\bar c_s$ would also be constant, and
the three terms in eq.~(\ref{eq:fatdisk}) would scale with time at
different rates:
the first two would be constant, but the third $\propto t^2$.
Thus the equation could not hold for all $t$.  
The solution is to let $\bar c_s\propto t$
so that the first and third terms balance, in which case
$\bar h\propto\rh$.

In short, a centrifugally supported stellar disk would not remain thin
but would evolve toward a quasi-spherical, pressure-supported
configuration as long as it is rapidly accreting from the ambient
black-hole disk.  Once $\bar h\sim\rh$, the viscous time becomes $\sim
(\alpha\Omega)^{-1}$ at most.  In fact, if the ``star'' fills its Hill
sphere, then the angular momentum of its outer parts is not conserved,
while its inner parts may transfer angular momentum outward by global
bar instabilities on their own dynamical timescale, which is
$\lesssim\Omega^{-1}$.

At this point, it appears that energy is more problematic than
angular momentum.
The estimate (\ref{eq:tcooldisk}) suggests that the cooling time
of a quasispherical configuration should scale as $t^2$, whereas
$\rh/\dot\rh\propto t$.  This would seem to imply that cooling cannot
keep pace with accretion, and therefore that the star should fill or
overflow its Hill sphere.  But the matter is more subtle than this
because of the peculiarities of $\Gamma\approx 4/3$ polytropes.

For a completely nonrotating star of fixed mass,
the contraction
timescale is the Kelvin-Helmholtz time obtained from
eqs. (\ref{eq:tKH}) and (\ref{eq:betaofM}); putting $R_*=\rh$,
\be\label{eq:tKHH}
t_{{\rm KH},H}\approx 53.0\,\apn^{-1/5}\krel^{19/20}\mu^{-4/5}
(\lum/\epn)^{3/20}
M_8^{-1/5}r_3^{13/20}\left(\frac{M_*}{M_{\rm iso}}\right)^{1/6}\yr\,,
\ee
which is somewhat shorter than the accumulation time
$t_{\rm iso}$ [eq.~(\ref{eq:tiso})].
(Henceforth we write $M_*$ rather than $M_H$ for the mass
within the Hill sphere.)
The critical difference between the two timescales is the
factor $\beta_*\propto M_*^{-1/2}$ in eqs.~(\ref{eq:BE}) \&
(\ref{eq:tKH}), which reflects the nearly zero binding energy of a
spherical high-mass star, so that only a small fraction
of the internal energy need be radiated in order to cause the
star to contract substantially.
This factor does not occur in the vertical contraction of a disk,
of course.

Furthermore, the star could contract even without radiative losses because
it accretes material with negative total energy.
It can be shown that when $\rh\gg H$, so that fluid
streamlines can be approximated by ballistic test-particle trajectories,
material accretes onto the Hill sphere with a negative
Jacobi constant per unit mass,
\begin{equation}\label{eq:Jacobi}
H_J\equiv \frac{1}{2}\mathbf{v}^2~-\frac{GM_*}{|\mathbf{r}-\mathbf{r}_*|}
~-\frac{3}{2}(\Omega\Delta r)^2\,,
\end{equation}
where $\mathbf{r}_*$ is the center of the star, 
$\Omega$ its orbital angular velocity, and $\mathbf{v}$
is the test-particle velocity relative to the star in a frame rotating
at $\Omega$.  (The Jacobi constant of the accreting streamlines
cannot be arbitrarily negative, since $H_J\ge -(9/2)(\Omega\rh)^2$ in order
to cross the inner or outer Lagrange point.)
Very close to the star, at $|\mathbf{r}-\mathbf{r}_*|\ll\rh$,
$H_J$ reduces to the usual kinetic-plus-potential energy in the potential
of an isolated mass $M_*$.  Thus, the putatively spherical star
accretes material that has negative total energy even without radiative
losses.  This would cause contraction on a timescale even shorter than
(\ref{eq:tKHH}) if rotation could be ignored altogether.  However,
we expect that the star will in fact rotate with a median angular
velocity scaling in proportion to $\Omega$ as it gains mass.  Consequently,
the star's rotational energy cannot be entirely neglected unless it
is smaller than the binding energy given by eq.~(\ref{eq:BE}), 
which is marginally unlikely since $\beta_*\sim 10^{-2}$ at $M_{\rm iso}$.

To recap, the growing star cannot cool fast enough to become a
disk but will be more nearly spherical and therefore able to shed
angular momentum quickly; on the other hand, the residual rotational
energy, though small compared to its gravitational energy, seems likely
to stabilize it against rapid contraction.  The interplay between
energy loss, angular-momentum loss, and radiation-pressure dominance
(so that $\Gamma\approx 4/3$) is so intricate that it will be difficult
to decide whether the star overflows or detaches from 
its Hill sphere without elaborate calculations that are beyond the
scope of this paper.

Again, it would be useful to have recourse to
numerical simulation.  We are not aware of any calculations
for disks supported largely by radiation pressure, but
\citet{Bate_etal03} and \citet{Lufkin_etal03}
have recently simulated three-dimensional accretion from
a protostellar disk onto embedded
planets ranging in mass from that of earth ($M_\oplus$)
to that of Jupiter ($M_J$).
Both studies place their planets at $r_J=5.2\au$ and follow the evolution
for $\sim 10^2$ orbits in a disk with $h/r=0.05$.
Bate \emph{et al.} use an eulerian finite-difference method, whereas
Lufkin \emph{et al.} use smooth particle hydrodynamics.  Neither includes
cooling, but this is not an impediment to accretion because 
the former study removes removes the gas
at each time step from the grid cells closest to the planet
(well within the Hill sphere), while the latter uses an isothermal
equation of state and includes the selfgravity of the gas, which allows
indefinite compression of the planet's atmosphere.
The accretion rates measured in these simulations are indeed
comparable, up to factors $\sim 2$,
to what is predicted by eq. (\ref{eq:tiso}) with $f_H=1$, even though
the most massive planets may exceed the isolation mass (eq. \ref{eq:Miso}),
$M_{\rm iso}= (0.089,0.25) \,f_H^{3/2}M_J$, corresponding to the
surface densities $\Sigma(r_J)\approx (0.23,0.46)\,M_J/r_J^2$ in the two
studies. (Bate \emph{et al.} impose a partially
cleared gap around the planet in their initial conditions; we have
quoted $\Sigma$ as it would be if there were no gap.)

Unfortunately for the purposes of this comparison,
the Hill radii of the most massive
planets in these simulations is scarcely larger than the disk thickness,
so that the accretion is not very far into the two-dimensional regime
contemplated by equation (\ref{eq:tiso}).
To express this another way, the planetary isolation mass quoted
above is comparable to the mass $M_h$ defined by $\rh(M_h)=h$,
\emph{viz.} $M_h=3(h/r)^3 M_\odot\approx0.38 M_J$.
The ratio $M_h/M_{\rm iso}$ scales as $(Qh/r)^{3/2}$, which is of order unity
for gaseous protoplanetary disks but very small at $r\sim 10^3\rs$
in quasar disks because both $h/r$ and $Q$ are smaller there.
Satellites $\ll M_h$ probably cannot open a gap regardless
of the viscosity of the disk \citep{Lin_Papaloizou93}; 
they will undergo Bondi accretion provided that the accumulated
gas cools rapidly enough.  Thus, the applicability of the concept of
isolation mass in quasar disks remains to be tested.

Once the isolation mass is achieved, or possibly even earlier, the
star will detach from its Hill sphere and contract until it reaches
the main sequence after a time of order the Kelvin time
(eq. \ref{eq:tkhms}).  It is important to note that, at least for our
nominal parameters, {\it the time required to reach the main sequence
is longer than the time to accrete the isolation mass}: therefore we
do not expect nuclear burning to impede accretion.  The formation time
is also short compared to the viscous evolution time of the disk
(eq. \ref{eq:tvisc}).

\subsection{Orbital migration and tidal disruption}

When fusion begins, the star may disintegrate \emph{via} pulsational
instability (\S\ref{sec:stability}).  It would certainly be interesting
to understand this stage: will the star disrupt completely or leave
behind a remnant?  How much will the disk be enriched?
Since the answers to these questions are not available at present,
we assume provisionally that the star survives the instability with
much of its mass intact.  If so, then it can be expected to undergo
Type II radial migration: that is, while maintaining an annular gap
in the disk by tidal torques, the star will drift inwards on the
viscous timescale (eq. \ref{eq:tvisc}) by analogy with jovian
planets in protostellar disks  \citep{Ward97}.  For our nominal
parameters, the latter timescale is shorter
than the main-sequence lifetime (eq. \ref{eq:tms}).  Thus
the migration may be completed before the star leaves the main
sequence.

The star will be tidally disrupted at an orbital radius where $\rh$ equals
the main-sequence radius (eq. \ref{eq:RofM}),
\begin{equation}\label{eq:rtidal}
r_{\rm tid} \approx 12\left(\frac{\mu_*}{\mu_\odot}\right)^{0.95}
\left(\frac{M_*}{10^5\msun}\right)^{0.14} M_8^{-2/3}\rs,
\end{equation}
which is comparable to the radius of the marginally stable orbit
for a Schwarzschild black hole with $M=10^9\msun$.

\subsection{Capture}

The concept of isolation mass is independent of the initial $Q$ of the
disk, although it is true that $M_{\rm iso}$ increases with the overall
mass of the disk [eq.~(\ref{eq:Miso})].  Thus mechanisms other than
gravitational instability may provide the initial seed from which the
isolation mass develops by accretion.  In the primordial solar nebula,
for example, where most likely $Q\sim10^2$ at the present position of
Jupiter, that planet is believed to have been initiated by collisional
agglomeration of solid bodies.  We do not expect solids in the non-self-gravitating
parts of quasar disks, but there may be other ways to form seeds that might
grow to the isolation mass.  One such process is capture of pre-existing stars
in collisions with the disk.

The capture
process has been examined at length by \citet{Syer_etal91} and
\citet{Artymowicz_etal93}, and we have only a few remarks to add.  In
the following discussion, all disk and black-hole parameters are
assumed equal to their nominal values in \S\ref{sec:initial} except
the orbital radius, for which we assume $r_3<1$ so that self-gravity
is slight and radiation pressure dominant.

The first remark is that in these regions, stars may perhaps more
often be destroyed by their encounters with the disk than captured.
The energy that must be dissipated to circularize a typical stellar
orbit exceeds the binding energy of a solar-type star by a factor
$\sim 10^3 r_3^{-2}$.  Most of the dissipation should occur in a bow
shock driven into the tenuous disk gas rather than within the star.
From \S\ref{subsec:physcon},
the peak postshock pressure is $\sim\rho v^2\sim 10^8
r_3^{-8/5}\,\mbox{dyne cm}^{-2}$; at just $\sim 10^{-9}$ of the star's
central pressure, this is only a gentle squeeze.  However, except
near the stagnation point, the postshock flow passes the star at
$\sim 10^4 r_3^{-1/3}\kms$ and at a temperature between $10^6
r_3^{-2/3}\K$ and $\sim 10^9 r_3^{-8/3}\K$ (depending whether the
radiation field comes to equilibrium with the plasma; the mean free
path for creation of soft photons by bremsstrahlung is comparable to
the standoff distance of the shock), so that it
may erode the stellar surface,
which is not rigid but simply a contact discontinuity.  Of
course, in view of the very low density of the disk gas ($\sim
10^{-9}r_3^{-3/5}\mbox{g cm}^{-3}$), the mass lost from the star in a
single disk passage is bound to be very small, but so is the orbital
energy; some $\Sigma R_\odot^2/M_\odot\sim 10^6 r_3^{3/5}$ passages
and $\sim 10^7 r_3^{21/10}\yr$ elapse before circularization.
A small minority of stars on orbits of low inclination
and low eccentricity may encounter the disk at less than their own
surface escape speed. Clearly the capture process presents further
opportunities for interesting hydrodynamical, or radiation-hydro,
simulations; \citet{Armitage_Zurek_Davies96} have simulated the passage
of giant stars through a disk and find that much of the envelope can be
stripped in a single encounter, but we are not aware of any hydrodynamic
simulations for main-sequence stars.

The second remark concerns the rate of accretion by the star once it
settles into the disk.  \citet{Syer_etal91} and \citet{Artymowicz_etal93}
estimate this at the Bondi rate,
\[
\dot M_{*,\rm B}= \frac{4\pi(GM_*)^2}{\cs^3}\approx 5\times10^{-3}
r_3^{21/10} \msun\yr^{-1},
\]
provided only that the Hill radius
is smaller than the disk thickness, \emph{i.e.} $M_*\lesssim
300 r_3^3 M_\odot$.  But accretion rate will be smaller if the gas
is unable to cool sufficiently quickly; in other words, 
the rate should not exceed Eddington rate,
\[
\dot M_{*,\rm E}= \frac{4\pi c}{\kappa} R_*\sim 1.0\times 10^{-3}
\frac{M_*/R_*}{\msun/r_\odot}\msun\yr^{-1}\,.
\]
Each of the first several doublings of the stellar mass is likely to
require $\sim 10^3\yr$, which may be gradual enough so that
a gap will have formed by the time $M_*\sim M_h$.

\section{Observable consequences}\label{sec:observe}

Exotic and luminous as stars of $\sim 10^5\msun$ would be, they could
easily have escaped attention.  Such stars radiate at their Eddington
luminosity, but that is smaller than the Eddington limit of the
quasar itself by the mass ratio $M_*/M\sim 10^{-3}$.  Their effective
temperature on the main sequence is $\approx 7\times 10^4\K$, based
on the radius (eq. \ref{eq:RofM}), which is determined essentially by
virial considerations;  a detailed atmospheric model might
give a somewhat different photospheric radius and temperature.
This is about twice the inferred $\Teff$ of bright QSOs
\citep{Malkan83}.

\begin{figure}
\epsscale{1.0}
\plotone{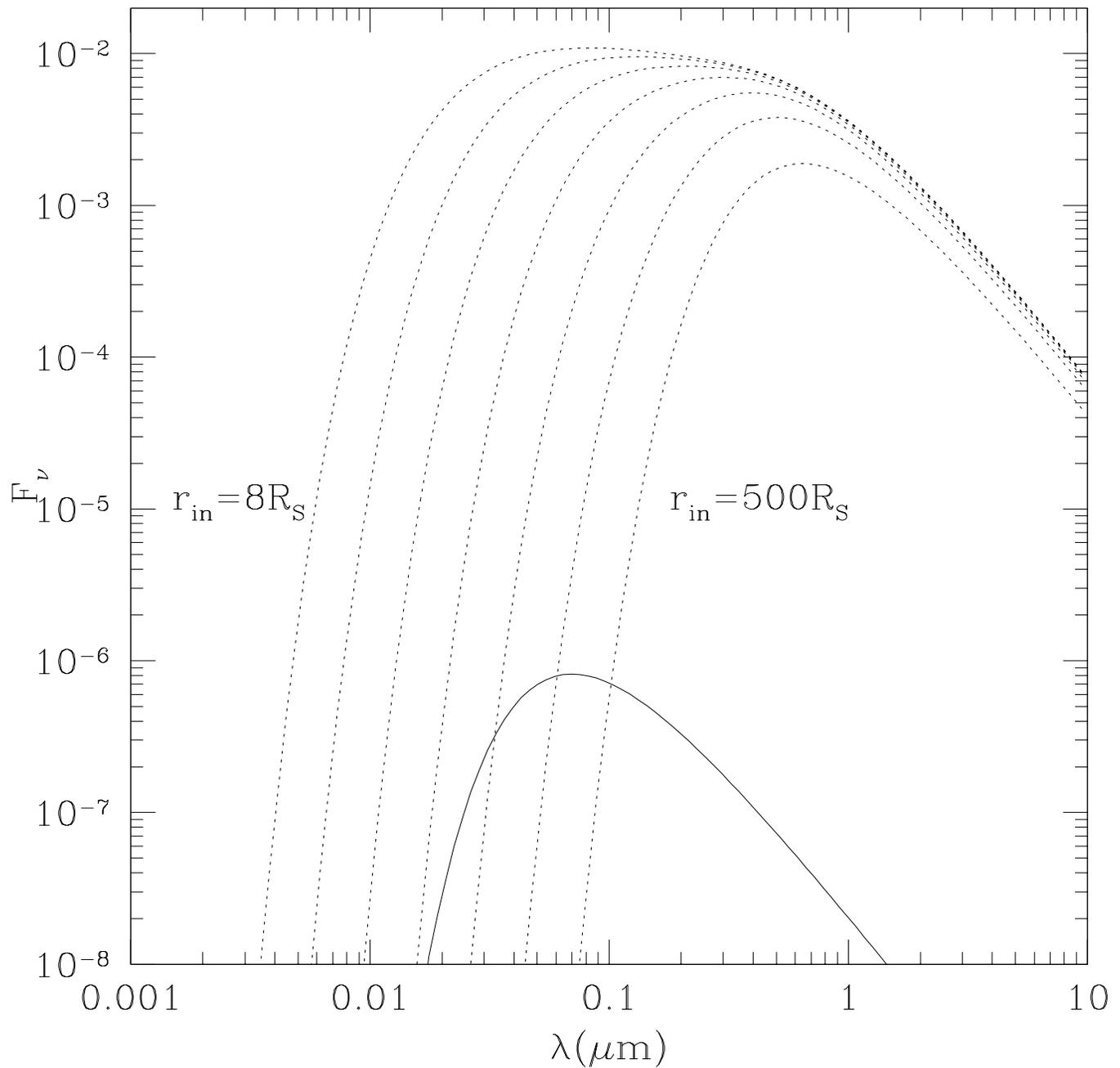}
\caption{
\label{fig:spec}
Evolution of $10^8\msun$ QSO accretion disk rest frame spectrum in
arbitrary units (dotted lines) as the inner radius of optically thick
emission migrates inwards from $500 R_S$ in steps of a factor of
two. The outer radius is taken to be $1000 R_S$
The spectrum (blackbody) from a main sequence, Eddington
luminosity supermassive star of mass equal to the isolation mass is
shown by the solid line.
}
\end{figure}

If the star has opened up a gap in the accretion disk, then this will
affect the quasar's spectrum, particularly when the gap is in the
innermost regions. In fact, because of the short viscous time of the
inner disk, we expect the gas inside the gap to be accreted relatively
quickly, leaving an evacuated region between the central black hole
and supermassive star, which is surrounded by what is effectively a
circumbinary disk.  The quasar will be relatively dim and red just
before disruption or accretion of the star:
indeed the source may not have been previously
identified as a quasar, given the migration time of $\sim 10^5\:{\rm
yr}$. This is the time when any supermassive star would be most easily
observed. For favorable inclinations, the star's spectral features
would exhibit very fast orbital velocities of order several thousand
$\kms$ with periods greater than several days. The emission from the
QSO disk, would gradually brighten and harden. Figure \ref{fig:spec}
shows an approximate comparison between the spectrum of the
supermassive star and QSO disk as the inner gas radius decreases.

\begin{figure}
\epsscale{0.9}
\plotone{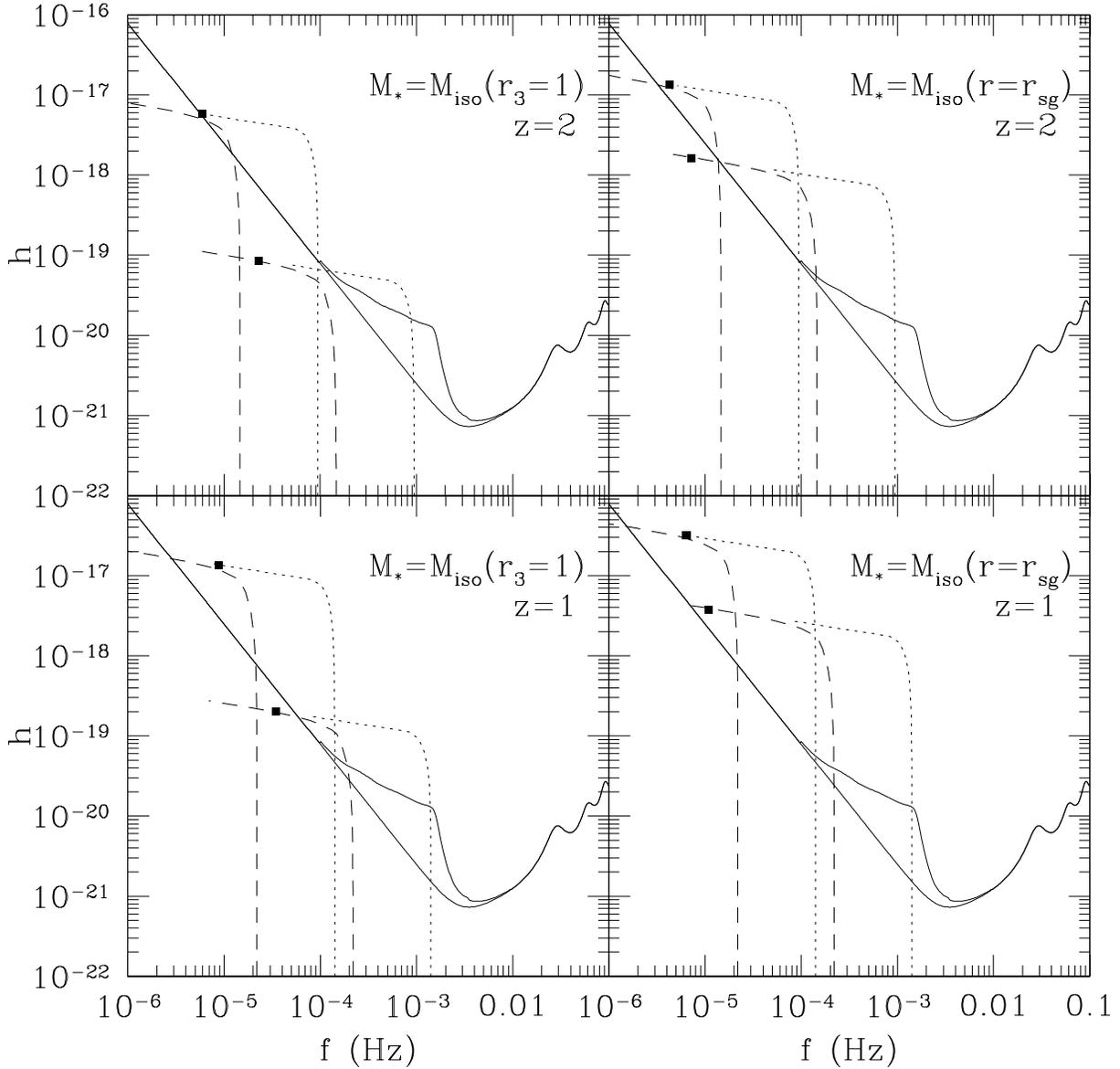}
\caption{
\label{fig:lisa}
\footnotesize
Gravitational waves from the inspiral of a compact object of mass
$M_{\rm iso}$ that formed at $r=10^3 R_S,r_{\rm s.g.}$ (left, right
panels), at redshift $z=1,2$ (bottom, top panels; standard
$\Omega_\Lambda=0.7$ cosmology assumed) from a supermassive black hole of
mass $10^7M_\odot$ (lower lines) and $10^8 M_\odot$ (upper lines),
as observed by LISA in a 1 year integration. For each case the ``modified
characteristic amplitude'' for a harmonic of the waves 
\citep{Finn_Thorne00}
from inspiral to a Schwarzschild black hole is shown by the
dashed line and to a Kerr black hole with rotation parameter of 0.999
by the dotted line. The square marks the point of disruption if the
secondary is a main sequence star. The lower solid line shows LISA's
rms instrumental noise level 
\citep{Larson_Hiscock00},
averaged over the sky with normalization as in \citet{Finn_Thorne00}, 
while the upper solid line is the total noise including an estimate for the
stochastic-background noise produced by white dwarf binaries 
\citep{Bender_Hils97}.
}
\end{figure}

The merger of a supermassive star (or supermassive remnant) with the
central black hole would be a strong source of gravitational waves.
We estimate the ``modified characteristic amplitude'' for quadrupole
gravitational waves as defined by \citet{Finn_Thorne00} from the
inspiral of a compact source with a mass equal to the isolation mass
formed at $10^3 R_s$ from the central black hole (Figure
\ref{fig:lisa}).  The trajectories in the frequency-amplitude diagram
depend on the angular momentum of the central black hole, since the
innermost stable orbit is much tighter for maximally rotating Kerr
black holes: in the extreme case shown in Figure \ref{fig:lisa} with
rotation parameter a=0.999, this radius is 1.18~$GM/c^2$, while it is
$6GM/c^2$ for the Schwarzschild black hole. For the systems shown in
Figure \ref{fig:lisa}, a main sequence supermassive star would be
disrupted before the differences due to angular momentum show
themselves.  Nevertheless, at and shortly after disruption, the
waveform is bound to differ from that expected
for an inspiraling black-hole companion.  It would be interesting
to try to predict these differences in some quantitative
detail, as they might be the most direct way to confirm the existence
of supermassive main-sequence stars in these disks.

The dependence of the isolation mass on the mass of
the black hole means that we
expect the supermassive stellar companions of more massive quasars to
be more readily detected, although the signal is shifted towards lower
frequencies. The cosmological redshift of the sources is important,
as illustrated in Figure \ref{fig:lisa}.
The evolution of the gravitational wave signal from inspiral, should also
be accompanied by a brightening and hardening of the quasar's
electromagnetic radiation, as described above, and may be marked by an
outburst if the star is disrupted.

We can make a crude estimate of the upper limit to the event rate of
supermassive star merger events from the known quasar population that
may be seen by LISA. Using the quasar luminosity function of
\citet{Boyle_etal00}, we estimate that the space density of
``typical'' quasars (within a couple of magnitudes of the knee of the
luminosity function) is $\sim 10^{-6}\:{\rm Mpc^{-3}}$ in a flat
$\Omega_\Lambda=0.7$ cosmology.  The comoving volume from $0.5<z<1.5$ is
$320\:{\rm Gpc^3}$ and from $1.5<z<2.5$ is $480\:{\rm Gpc^3}$ so there
are $2.3, 3.5 \times 10^5$ typical QSOs in these intervals,
respectively, that can be seen over the whole sky.  The supermassive
star formation rate cannot be much greater than one per viscous time
at $r_{\rm s.g.}$ (and could be substantially less), which is about
$1\times 10^{-5}\:{\rm yr^{-1}}$ per quasar. Thus the maximum event
rate for LISA is of order a few supermassive star - black hole mergers
per year. Of course the observed quasar luminosity function may
underestimate the true number because of beaming effects.

\section{Conclusions}\label{sec:conclusions}

We have argued that conditions in QSO accretion disks are likely to
lead to the formation of massive stars. If the final masses of these
stars are limited by the opening of a gap in the disk near the
innermost radius where the disk is self-gravitating, then they would
have $\sim 10^5\sm$.  This is much more massive than any stellar object
currently known. A key point is that the main sequence lifetime of
these stars is much longer than their formation time, and
modestly longer than the time they
take to migrate to the black hole. Thus the end point of stellar
evolution does not prevent the existence of these supermassive stars,
and indeed the evolution may be interrupted only by tidal disruption and
merger near the central black hole.  A second important point is that
once a star has formed in the disk with initial mass of order the
Toomre mass, $\sim 100\sm$, then the subsequent accretion rates of
surrounding disk material are much larger ($\sim \smyr$) than the mass
loss rates caused by thermonuclear instabilities ($< 10^{-3}\smyr$).

Radiative feedback from the forming massive star (from luminosity
generated either by nuclear burning or gravitational contraction) will
have little influence while conditions in the disk and accretion flow
are optically thick. Gas streamlines will only be significantly
perturbed as the stellar surroundings become optically thin, probably
close to the time of gap formation, when the isolation mass has
already been reached.


Though they are very luminous, supermassive stars would easily be lost
in the glare of the quasar.  Periodic modulations of the light curve
due to such stars would have periods ranging from a few years to a few
days but with millimagnitude amplitudes.  Even if the quasar was in a
quiet state, perhaps because of the disruption of its disk by the
supermassive star, then the light from the central regions of the host
galaxy might dominate.  The best candidates for periodicity searches
would be quasars that appear to lack a blue bump, as this might be due
to a gap opened in the disk by such a star.  Space-based gravitational
wave detectors are promising tools in the search for supermassive
stars, particularly since the signal should also be accompanied by a
rapid brightening of the quasar as the star is disrupted.

Very recently, ideas similar to those presented here have been
presented by Levin (2003).  Although we have discussed
our respective drafts, our two efforts have been independent.
There is much we agree on, but there are some differences in
assumptions and emphasis.  Whereas Levin considers a disk
that fragments into many stars of up to a few hundred solar
masses,
we consider the formation of a single dominant mass
that accretes a significant fraction of the entire disk.
Naturally we prefer our own scenario, but much more work
will be required to decide which (if either) is correct.
It is also possible that the outcome depends upon the initial
conditions of the disk before fragmentation begins.  Some insight
into the early stages of fragmentation
could be gained by numerical simulations of marginally self-gravitating
disks; there have been many such studies, but none that we are aware of
that include radiation-pressure dominance and radiative diffusion.
Another difference between our studies is that and Levin considers
parameters appropriate to the Galactic Center, whereas we focus on
bright and massive QSOs.  \cite{Collin_Zahn99a} and \cite{Collin_Zahn99b}
have also discussed star formation in AGN disks recently, but they too
consider stars which, though massive by ordinary standards, are much less
than the isolation mass considered here.

Another area of theoretical uncertainty, which is perhaps even more
challenging, is the nonlinear stability and evolution of extremely
massive stars.  In recent years, apart from some important
contributions cited above, work on this topic has languished for lack
of a clear astrophysical motivation.  Convincing observational
evidence for stars above $\sim 10^2 M_\odot$ does not exist.  But, as
we have already stressed, the physical conditions in a quasar
accretion disk are very much more extreme than those of a conventional
star-forming region, and we have argued that the self-gravity and
short timescales of these disks favor the rapid assembly of truly
supermassive protostars.  It would be very interesting to revisit the
question of whether such objects can survive on the main sequence.

\acknowledgments

We thank Yuri Levin and two anonymous referees for helpful discussions. JCT
is supported by a Spitzer-Cotsen fellowship from Princeton University
and by NASA grant NAG 5-10811.  This work was supported in part by NSF
grant AST-0307558 to JG.

\bibliography{apj,agn,disk,gal,misc,mri,planet,stars}

\begin{thebibliography}{}

\bibitem[\protect\citeauthoryear{{Appenzeller}}{{Appenzeller}}{1970a}]{Appenze%
ller70b}
{Appenzeller} I.,  1970a, A\&A, 9, 216

\bibitem[\protect\citeauthoryear{{Appenzeller}}{{Appenzeller}}{1970b}]{Appenze%
ller70a}
{Appenzeller} I.,  1970b, A\&A, 5, 355

\bibitem[\protect\citeauthoryear{{Armitage}, {Zurek} \& {Davies}}{{Armitage}
  et~al.}{1996}]{Armitage_Zurek_Davies96}
{Armitage} P.~J.,  {Zurek} W.~H.,    {Davies} M.~B.,  1996, ApJ, 470, 237

\bibitem[\protect\citeauthoryear{{Artymowicz}, {Lin} \& {Wampler}}{{Artymowicz}
  et~al.}{1993}]{Artymowicz_etal93}
{Artymowicz} P.,  {Lin} D.~N.~C.,    {Wampler} E.~J.,  1993, ApJ, 409, 592

\bibitem[\protect\citeauthoryear{Bate, Lubow, Ogilvie \& Miller}{Bate
  et~al.}{2003}]{Bate_etal03}
Bate M.~R.,  Lubow S.~H.,  Ogilvie G.~I.,    Miller K.~A.,  2003,
  Three-dimensional calculations of high and low-mass planets embedded in
  protoplanetary discs, {\tt astro-ph/0301154}

\bibitem[\protect\citeauthoryear{{Baumgarte} \& {Shapiro}}{{Baumgarte} \&
  {Shapiro}}{1999}]{Baumgarte_Shapiro99}
{Baumgarte} T.~W.,  {Shapiro} S.~L.,  1999, ApJ, 526, 941

\bibitem[\protect\citeauthoryear{Bender \& Hils}{Bender \&
  Hils}{1997}]{Bender_Hils97}
Bender P.~L.,  Hils D.,  1997, {Classical and Quantum Gravity}, 14, 1439

\bibitem[\protect\citeauthoryear{{Bond}, {Arnett} \& {Carr}}{{Bond}
  et~al.}{1984}]{Bond_Arnett_Carr84}
{Bond} J.~R.,  {Arnett} W.~D.,    {Carr} B.~J.,  1984, ApJ, 280, 825

\bibitem[\protect\citeauthoryear{{Boyle}, {Shanks}, {Croom}, {Smith}, {Miller},
  {Loaring} \& {Heymans}}{{Boyle} et~al.}{2000}]{Boyle_etal00}
{Boyle} B.~J.,  {Shanks} T.,  {Croom} S.~M.,  {Smith} R.~J.,  {Miller} L.,
  {Loaring} N.,    {Heymans} C.,  2000, MNRAS, 317, 1014

\bibitem[\protect\citeauthoryear{{Chandrasekhar}}{{Chandrasekhar}}{1964}]{Chan%
dra64}
{Chandrasekhar} S.,  1964, ApJ, 140, 417

\bibitem[\protect\citeauthoryear{{Chokshi} \& {Turner}}{{Chokshi} \&
  {Turner}}{1992}]{Chokshi_Turner92}
{Chokshi} A.,  {Turner} E.~L.,  1992, MNRAS, 259, 421

\bibitem[\protect\citeauthoryear{{Collin} \& {Zahn}}{{Collin} \&
  {Zahn}}{1999a}]{Collin_Zahn99a}
{Collin} S.,  {Zahn} J.,  1999a, A\&A, 344, 433

\bibitem[\protect\citeauthoryear{{Collin} \& {Zahn}}{{Collin} \&
  {Zahn}}{1999b}]{Collin_Zahn99b}
{Collin} S.,  {Zahn} J.,  1999b, Ap\&SS, 265, 501

\bibitem[\protect\citeauthoryear{{Dietrich}, {Hamann}, {Shields}, {Constantin},
  {Heidt}, {J{\" a}ger}, {Vestergaard} \& {Wagner}}{{Dietrich}
  et~al.}{2003}]{Dietrich03}
{Dietrich} M.,  {Hamann} F.,  {Shields} J.~C.,  {Constantin} A.,  {Heidt} J.,
  {J{\" a}ger} K.,  {Vestergaard} M.,    {Wagner} S.~J.,  2003, ApJ, 589, 722

\bibitem[\protect\citeauthoryear{Eddington}{Eddington}{1926}]{Eddington26}
Eddington A.~S.,  1926, The Internal Constitution of the Stars.
Cambridge University Press

\bibitem[\protect\citeauthoryear{{Finn} \& {Thorne}}{{Finn} \&
  {Thorne}}{2000}]{Finn_Thorne00}
{Finn} L.~S.,  {Thorne} K.~S.,  2000, Phys. Rev. D, 62, 124021

\bibitem[\protect\citeauthoryear{{Fryer}, {Woosley} \& {Heger}}{{Fryer}
  et~al.}{2001}]{Fryer_Woosley_Heger01}
{Fryer} C.~L.,  {Woosley} S.~E.,    {Heger} A.,  2001, ApJ, 550, 372

\bibitem[\protect\citeauthoryear{{Gammie}}{{Gammie}}{2001}]{Gammie01}
{Gammie} C.~F.,  2001, ApJ, 553, 174

\bibitem[\protect\citeauthoryear{Goodman}{Goodman}{2003}]{Goodman03}
Goodman J.,  2003, MNRAS, 339, 937

\bibitem[\protect\citeauthoryear{{Hoyle} \& {Fowler}}{{Hoyle} \&
  {Fowler}}{1963}]{Hoyle_Fowler63b}
{Hoyle} F.,  {Fowler} W.~A.,  1963, MNRAS, 125, 169

\bibitem[\protect\citeauthoryear{{Kumar}}{{Kumar}}{1999}]{Kumar99}
{Kumar} P.,  1999, ApJ, 519, 599

\bibitem[\protect\citeauthoryear{{Larson}, {Hiscock} \& {Hellings}}{{Larson}
  et~al.}{2000}]{Larson_Hiscock00}
{Larson} S.~L.,  {Hiscock} W.~A.,    {Hellings} R.~W.,  2000, Phys. Rev. D, 62,
  62001

\bibitem[\protect\citeauthoryear{Ledoux}{Ledoux}{1941}]{Ledoux41}
Ledoux P.,  1941, ApJ, 94, 537

\bibitem[\protect\citeauthoryear{Levin}{Levin}{tted}]{Levin03}
Levin Y.,  submitted, Formation of massive stars and black holes in
  self-gravitating AGN discs, {\tt astro-ph/0307084}

\bibitem[\protect\citeauthoryear{{Lin} \& {Papaloizou}}{{Lin} \&
  {Papaloizou}}{1993}]{Lin_Papaloizou93}
{Lin} D.~N.~C.,  {Papaloizou} J.~C.~B.,  1993, in Protostars and Planets III
  {On the tidal interaction between protostellar disks and companions}.
pp 749--835

\bibitem[\protect\citeauthoryear{Lissauer}{Lissauer}{1987}]{Lissauer87}
Lissauer J.~J.,  1987, Icarus, 69, 248

\bibitem[\protect\citeauthoryear{Lufkin, Quinn, Wadsley, Stadel \&
  Governato}{Lufkin et~al.}{2003}]{Lufkin_etal03}
Lufkin G.,  Quinn T.,  Wadsley J.,  Stadel J.,    Governato F.,  2003,
  {Simulations of Gaseous Disc-Embedded Planet Interaction}, {\tt
  astro-ph/0305546}

\bibitem[\protect\citeauthoryear{{Malkan}}{{Malkan}}{1983}]{Malkan83}
{Malkan} M.~A.,  1983, ApJ, 268, 582

\bibitem[\protect\citeauthoryear{{McKee} \& {Tan}}{{McKee} \&
  {Tan}}{2003}]{McKee_Tan03}
{McKee} C.~F.,  {Tan} J.~C.,  2003, ApJ, 585, 850

\bibitem[\protect\citeauthoryear{{Paczy{\' n}ski}}{{Paczy{\' n}ski}}{1977}]{Paczynski77}
{Paczy{\' n}ski} B.,  1977, ApJ, 216, 822

\bibitem[\protect\citeauthoryear{{Papaloizou}}{{Papaloizou}}{1973}]{Papaloizou%
73}
{Papaloizou} J.~C.~B.,  1973, MNRAS, 162, 169

\bibitem[\protect\citeauthoryear{{Plume}, {Jaffe}, {Evans}, {Martin-Pintado} \&
  {Gomez-Gonzalez}}{{Plume} et~al.}{1997}]{Plume_etal97}
{Plume} R.,  {Jaffe} D.~T.,  {Evans} N.~J.,  {Martin-Pintado} J.,
  {Gomez-Gonzalez} J.,  1997, ApJ, 476, 730

\bibitem[\protect\citeauthoryear{{Pringle}}{{Pringle}}{1981}]{Pringle81}
{Pringle} J.~E.,  1981, ARAA, 19, 137

\bibitem[\protect\citeauthoryear{{Rafikov}}{{Rafikov}}{2003}]{Rafikov03c}
{Rafikov} R.~R.,  2003, AJ, 125, 942

\bibitem[\protect\citeauthoryear{{Rees}}{{Rees}}{1984}]{Rees84_ARAA}
{Rees} M.~J.,  1984, ARAA, 22, 471

\bibitem[\protect\citeauthoryear{{Schaller}, {Schaerer}, {Meynet} \&
  {Maeder}}{{Schaller} et~al.}{1992}]{Schaller_etal92}
{Schaller} G.,  {Schaerer} D.,  {Meynet} G.,    {Maeder} A.,  1992, A\&AS, 96,
  269

\bibitem[\protect\citeauthoryear{Schwarzschild \& H\"arm}{Schwarzschild \&
  H\"arm}{1959}]{Schwarzschild_Harm59}
Schwarzschild M.,  H\"arm R.,  1959, ApJ, 129, 637

\bibitem[\protect\citeauthoryear{{Shakura} \& {Sunyaev}}{{Shakura} \&
  {Sunyaev}}{1973}]{Shakura_Sunyaev73}
{Shakura} N.~I.,  {Sunyaev} R.~A.,  1973, A\&A, 24, 337

\bibitem[\protect\citeauthoryear{{Shlosman} \& {Begelman}}{{Shlosman} \&
  {Begelman}}{1987}]{Shlosman_Begelman87}
{Shlosman} I.,  {Begelman} M.~C.,  1987, Nature, 329, 810

\bibitem[\protect\citeauthoryear{{Shlosman} \& {Begelman}}{{Shlosman} \&
  {Begelman}}{1989}]{Shlosman_Begelman89}
{Shlosman} I.,  {Begelman} M.~C.,  1989, ApJ, 341, 685

\bibitem[\protect\citeauthoryear{{Shlosman}, {Begelman} \& {Frank}}{{Shlosman}
  et~al.}{1990}]{Shlosman_Begelman_Frank90}
{Shlosman} I.,  {Begelman} M.~C.,    {Frank} J.,  1990, Nature, 345, 679

\bibitem[\protect\citeauthoryear{{Sirko} \& {Goodman}}{{Sirko} \&
  {Goodman}}{2003}]{Sirko_Goodman03}
{Sirko} E.,  {Goodman} J.,  2003, MNRAS, 341, 501

\bibitem[\protect\citeauthoryear{{Soltan}}{{Soltan}}{1982}]{Soltan82}
{Soltan} A.,  1982, MNRAS, 200, 115

\bibitem[\protect\citeauthoryear{{Syer}, {Clarke} \& {Rees}}{{Syer}
  et~al.}{1991}]{Syer_etal91}
{Syer} D.,  {Clarke} C.~J.,    {Rees} M.~J.,  1991, MNRAS, 250, 505

\bibitem[\protect\citeauthoryear{{Toomre}}{{Toomre}}{1964}]{Toomre64}
{Toomre} A.,  1964, ApJ, 139, 1217

\bibitem[\protect\citeauthoryear{{Vink}, {de Koter} \& {Lamers}}{{Vink}
  et~al.}{2000}]{Vink_etal00}
{Vink} J.~S.,  {de Koter} A.,    {Lamers} H.~J.~G.~L.~M.,  2000, A\&A, 362, 295

\bibitem[\protect\citeauthoryear{{Ward}}{{Ward}}{1997}]{Ward97}
{Ward} W.~R.,  1997, Icarus, 126, 261

\bibitem[\protect\citeauthoryear{{Yu} \& {Tremaine}}{{Yu} \&
  {Tremaine}}{2002}]{Yu_Tremaine02}
{Yu} Q.,  {Tremaine} S.,  2002, MNRAS, 335, 965

\bibitem[\protect\citeauthoryear{{Zel'dovich} \& Novikov}{{Zel'dovich} \&
  Novikov}{1971}]{Zeldovich_Novikov71}
{Zel'dovich} Y.~B.,  Novikov I.~D.,  1971, Relativistic Astrophysics, Volume 1,
  Stars and Relativity.
University of Chicago Press

\end{thebibliography}
\bibliographystyle{mn2e}

\end{document}